\documentstyle[12pt]{article}

 \def\ex{{\hbox{\rm e}}}

  \def\tr{{\hbox{\rm Tr}}}

 \def\vev{vacuum expectation value}

\def\ie{{\em i.e.}}

\newcommand{\ZZ}{{\mbox{{\bf Z}}}}
\newcommand{\RR}{{\mbox{{\bf R}}}}
\newcommand{\RRs}{{\mbox{{\small \bf R}}}}
\newcommand{\CS}{{\scriptstyle {\rm CS}}}
\newcommand{\CSs}{{\scriptscriptstyle {\rm CS}}}

\csname @addtoreset\endcsname{equation}{section}

\begin{document}
\begin{titlepage}
\begin{flushright}
USC-FT-26-95 \\
hep-th/9511037
\end{flushright}
\vglue 1cm

{}~\vfill

\begin{center}

{\large \bf  TOPOLOGICAL QUANTUM FIELD THEORY: \protect \\
{}~\vfill
A  PROGRESS REPORT\footnotemark}

\footnotetext{Talk given at the IV Fall Workshop on Differential Geometry
and its Applications, Santiago de Compostela, 1995}

\vskip5cm

{\large J. M. F. Labastida\footnotemark}

\footnotetext{e-mail: labastida@gaes.usc.es}

\vspace{1cm}

{\em Departamento de F\'\i sica de Part\'\i culas \\
Universidade de Santiago de Compostela \\
E-15706 Santiago de Compostela, Spain}

\end{center}

\vskip3cm

\begin{abstract}
A brief introduction to Topological Quantum
Field Theory as well as a description of recent progress made in the
field is presented. I concentrate mainly on the connection between
Chern-Simons gauge theory and Vassiliev invariants, and Donaldson theory and
its
generalizations and Seiberg-Witten invariants. Emphasis is made on the
usefulness of these relations to obtain explicit expressions for topological
invariants, and on the universal structure underlying both systems.
\end{abstract}

\vskip2cm


\end{titlepage}

\pagebreak

\section{Introduction}

During the last decade we have witnessed a striking new relation between
mathematics and physics. This relation connects many of the most
advanced ideas in the two fields involved, topology and quantum physics.
On the one side, the most sophisticated topological invariants of three and
four-dimensional manifolds are encountered. On the other side, the most recent
achievements in quantum field theory play a salient role.
It is remarkable to observe that precisely these low dimensions in which
topology has shown to present important features are the dimensions where
many interesting quantum field theories are renormalizable.

Though connections between quantum physics and topology can be traced back
to the fifties, it is in the eighties when a new and unprecedented kind of
relation between the two takes place. In 1982 E. Witten \cite{morse}
considered $N=2$ supersymmetric
sigma models in two dimensions and rewrote Morse theory in the language of
quantum field theory. Furthermore, he constructed out of those models a refined
version of Morse theory known nowadays as Morse-Witten theory. Witten's
arguments in \cite{morse} made use of functional integrals and therefore can be
regarded as non-rigorous. Nevertheless, some years later A. Floer reformulated
Morse-Witten theory providing a rigorous mathematical structure
 \cite{floer}. This trend in which
some mathematical structure is first constructed  by quantum field theory
methods and then reformulated in a rigorous mathematical ground constitutes one
of the tendencies in this new relation between topology and physics.

The influence of M. Atiyah
\cite{atiyah} on E. Witten in the fall of 1987 culminated with the construction
by the latter of the first topological quantum field theory (TQFT) in January
1988  \cite{tqft}. The quantum theory turned out
to be a ``twisted" version of $N=2$ supersymmetric Yang-Mills. This theory,
whose
existence was conjectured by M. Atiyah, is related to Donaldson invariants for
four-manifolds    \cite{donald}, and it is known
nowadays as Donaldson-Witten theory.

In 1988 E. Witten formulated also two models which have been of fundamental
importance in two and three dimensions: topological sigma models
\cite{tsm} and Chern-Simons gauge theory  \cite{csgt}. The first
one can be understood as a twist of the $N=2$ supersymmetric sigma model
considered by Witten in his work on Morse theory \cite{morse}, and is related
to
Gromov invariants \cite{gromov}. The second one
is not the result of a twist but a model whose action is the integral of the
Chern-Simons form. In this case the corresponding topological invariants are
knot and link invariants as the Jones polynomial  \cite{jones} and its
generalizations.

TQFT provided a new point of view to study the
topological invariants which were discovered only a few years before the
formulation of this type of quantum theory. One of the important aspects of
this
new approach is that they could be generalized in a variety of directions.
Since
1988 there are two main lines of work: on the one hand, the rigorous
constructions (without using functional integration) of
the generalizations predicted by TQFT;
on the other hand, the use of quantum field theory techniques to analyze and
compute the generalized invariants. Both lines of work have provided a very
valuable outcome but, perhaps,  the most striking
results have been achieved by the
quantum field theory line of thought.

TQFTs have been studied from both, perturbative
and non-perturbative approaches. In physical theories it is well known that
both
approaches provide very valuable information on the features of the models
under consideration. In general, non-perturbative methods are less developed
than  perturbative ones. However, precisely in $N=2$ supersymmetric
theories, the ones intimately related to TQFTs,
important progress has been done recently
 \cite{sw}. It is also important to notice
that TQFTs are in general much simpler than their
physical counterparts and one expects that the use of these methods is much
more
tractable.

In three dimensions, non-perturbative methods have been applied to Chern-Simons
gauge theory to obtain properties of knot and link invariants as well as
general procedures for their computation. On the other hand, perturbative
methods have provided an integral representation for Vassiliev invariants
 \cite{vass}
which, among other things, allows to extend the formulation of these invariants
to arbitrary smooth three-manifolds. Vassiliev invariants are strong candidates
to classify knots and links and they have been much studied during the last
years.

In four dimensions, perturbative methods show that Donaldson-Witten theory is
related to Donaldson invariants. On the other hand, non-perturbative methods
indicate that those invariants are related to other rather different
topological
invariants which are called Seiberg-Witten invariants. In sharp contrast to
Donaldson invariants, which are defined on the moduli spaces of instantons,
Seiberg-Witten invariants are associated to moduli spaces of abelian monopoles
\cite{abm,park}. Recently, Donaldson-Witten theory has been generalized
to a theory
involving non-abelian monopoles
\cite{nabm}. This provides a rich set of
new topological invariants which opens new perspectives in the study of
four-manifolds. Nevertheless, there are indications that these new
invariants can
also be written, at least in some situations, in terms of Seiberg-Witten
invariants. Therefore, it might happen that no new topological information is
gained.

The situation for three and  four-dimensional TQFTs
is shown in Table \ref{latabla}. These theories  seem to share a common
structure. Their topological invariants are rich and can be labeled with group
theoretical data: Wilson lines for different representations and  gauge groups
(Jones polynomial and its generalizations) and non-abelian monopoles for
different representations and gauge groups (generalized Donaldson polynomials).
However, these invariants can all be written in terms of topological invariants
which are independent of the group and representations chosen: Vassiliev
invariants and Seiberg-Witten invariants, respectively. Both depend strictly on
the topology. The group-theoretical data labeling
generalized Jones and Donaldson polynomials enter in the coefficients of the
expressions of these polynomials as a power series in Vassiliev and
Seiberg-Witten invariants, respectively.

\begin{table}[htbp]
\centering
\begin{tabular}{|c|c|c|}  \hline
    & $d=3$  &  $d=4$ \\  \hline
perturbative & Vassiliev & Donaldson \\ \hline
non-perturbative & Jones & Seiberg-Witten \\ \hline
\end{tabular}
\caption{Topological invariants in the perturbative and the non-perturbative
regimes for $d=3$ and $d=4$.}
\label{latabla}
\end{table}

The resemblance between the two pictures is very appealing. Nevertheless, there
are important differences which rise some important questions. In the case of
knot theory, Vassiliev invariants constitute an infinite set. However, in
Donaldson theory, for the cases studied so far, only a finite set of invariants
seems to play a relevant role. One would like to know if this is general or if
this fact is just a peculiarity of the only two cases (gauge group $SU(2)$
without matter and with one multiplet of matter in the fundamental
representation) which have been studied so far. The general picture of
non-perturbative $N=2$ supersymmetric Yang-Mills theories seems to suggest that
the set of invariants entering the expressions for the generalized Donaldson
polynomials is going to be finite. However, one might find unexpected results
by studying different kinds of matter. Much work in this direction has to be
done.

In this talk I will concentrate mainly on a description of some of
the recent progress
made in the field. I will begin by introducing the framework of TQFTs in sec. 2
and their methods of construction in sec. 3. Excellent reviews on these aspects
are available \cite{thompson}, \cite{moore}, and I refer the reader to those
for
details. In sec. 4 I describe  the progress made in the last years in three
dimensions while in sec. 5 I do the same in four dimensions.
Two-dimensional TQFT, a very active subject during the last few years,
lie outside the scope of this talk. Finally,  some concluding remarks  are
presented in sec. 6.

\section{Topological Quantum Field Theory}

In this section we present the most general structure of a TQFT
from a functional
integral point of view. As in ordinary quantum field theory, the functional
integration involved is not in general well defined. Similarly to the case of
ordinary quantum field theory this has led to the construction of an axiomatic
approach
\cite{axio}. In this section, however, we are not going to
describe this approach. We will concentrate on the functional integral
point of view. Although not well defined in general, this is the approach which
more success has gathered.

Our basic topological space will be an $n$-dimensional Riemannian manifold $M$
endowed with a metric $g_{\mu\nu}$. Let us consider on it a set of
fields $\{\phi_i\}$, and let $S(\phi_i)$ be a real functional of
these fields which will be regarded as the action of the theory. We will
consider
``operators",
${\cal O}_\alpha(\phi_i)$, which are in general arbitrary functionals of the
fields. In TQFT these functionals are real
functionals labeled by some set of indices $\alpha$ carrying topological or
group-theoretical data. The vacuum expectation value (vev) of a product of
these
operators is defined as:
\begin{equation}
\langle {\cal O}_{\alpha_1} {\cal O}_{\alpha_2} \cdots
{\cal O}_{\alpha_p} \rangle =
\int [D\phi_i]  {\cal O}_{\alpha_1}(\phi_i) {\cal O}_{\alpha_2}(\phi_i)
 \cdots {\cal O}_{\alpha_p}(\phi_i) \exp\big(-S(\phi_i)\big).
\label{sanse}
\end{equation}
A quantum field theory is considered topological if the following relation is
satisfied:
\begin{equation}
\frac{\delta}{\delta g^{\mu\nu}}
\langle {\cal O}_{\alpha_1} {\cal O}_{\alpha_2} \cdots
{\cal O}_{\alpha_p} \rangle = 0,
\label{reme}
\end{equation}
\ie, if the vevs of some set of selected operators is
independent of the metric $g_{\mu\nu}$ on $M$. If such is the case those
operators are called ``observables".

There are two ways to guarantee, at least formally, that condition (\ref{reme})
is satisfied. The first one corresponds to the situation in which both, the
action $S(\phi_i)$, as well as the operators ${\cal O}_{\alpha_i}$ are metric
independent. These TQFTs are called of Schwarz
type. The most important representative is Chern-Simons gauge theory. The
second one corresponds to the case in which there exist a symmetry, whose
infinitesimal form will be denoted by $\delta$, satisfying the following
properties:
\begin{equation}
\delta {\cal O}_{\alpha_i} = 0,
\;\;\;\;\;
T_{\mu\nu} = \delta G_{\mu\nu},
\label{angela}
\end{equation}
where $T_{\mu\nu}$ is the energy-momentum tensor of the theory, \ie,
\begin{equation}
T_{\mu\nu} (\phi_i) = \frac{\delta}{\delta g^{\mu\nu}} S(\phi_i).
\label{rosi}
\end{equation}

The fact that $\delta$ in (\ref{angela}) is a symmetry of the theory implies
that the transformations $\delta\phi_i$ of the fields are such that both
$\delta S(\phi_i) =0 $ and $\delta {\cal O}_{\alpha_i} (\phi_i) =0 $.
Conditions
(\ref{angela}) lead, at least formally, to the following relation for vevs:
\begin{eqnarray}
& & \frac{\delta}{\delta g^{\mu\nu}}
\langle {\cal O}_{\alpha_1} {\cal O}_{\alpha_2} \cdots
{\cal O}_{\alpha_p} \rangle   = -\int [D\phi_i]
{\cal O}_{\alpha_1}(\phi_i) {\cal O}_{\alpha_2}(\phi_i)
 \cdots {\cal O}_{\alpha_p}(\phi_i) T_{\mu\nu} \exp\big(-S(\phi_i)\big)
\nonumber \\
&=& -\int [D\phi_i] \delta\Big(
{\cal O}_{\alpha_1}(\phi_i) {\cal O}_{\alpha_2}(\phi_i)
 \cdots {\cal O}_{\alpha_p}(\phi_i) G_{\mu\nu} \exp\big(-S(\phi_i)\big)\Big)
 =  0,
\label{rosina}
\end{eqnarray}
which implies that the quantum field theory can be regarded as topological.
This second type of TQFTs are called of Witten type.
One of its main representatives is the theory related to Donaldson invariants,
which, as already indicated, is a twisted version of $N=2$ supersymmetric
Yang-Mills theory. It is important to remark that the symmetry $\delta$ must be
a scalar symmetry, \ie, that its symmetry parameter must be a scalar. The
reason is that, being a global symmetry, this parameter must be covariantly
constant and for arbitrary manifolds this property, if it is satisfied at all,
implies strong restrictions unless the parameter is a scalar.

\section{Construction Methods}

As indicated in the introduction, the first TQFT
was constructed performing a twist on $N=2$ supersymmetric Yang-Mills. Since
then other construction methods have been used. These methods have
been useful because, on the one hand, they have led to the construction of new
TQFTs, and, on the other hand they have provided a
better geometric understanding of the meaning of the TQFT under
consideration. In
the first group stand out the method of twisting $N=2$ supersymmetry, as well
as methods based on the ``BRST" formalism also called cohomological approach.
In
the second group the Mathai-Quillen formalism is the most successful one.
Nowadays it can be considered as the best approach although often the
information learned from the other approaches turns out to be very valuable.
In this section I will
describe very briefly some of these approaches.

\vskip0.3cm
\noindent{\sc 3.1 Twisting $N=2$ supersymmetry}
\vskip0.1cm

\noindent
In this approach the starting point is an $N=2$ supersymmetric quantum
field theory. The basic ingredient of this method consists of extracting a
scalar
symmetry out of the $N=2$ supersymmetry. Let us consider the case of $d=4$. In
${\RR}^4$ the global symmetry group when $N=2$ supersymmetry is present is
$H= SU(2)_L\otimes SU(2)_R \otimes SU(2)_I \otimes U(1)_{\cal R} $ where
${\cal K} = SU(2)_L \otimes SU(2)_R$ is the rotation group and
$SU(2)_I \otimes U(1)_{\cal R}$ is the internal symmetry group. The
supercharges $Q^i_\alpha$ and $\overline Q_{i\dot\alpha }$ which generate
$N=2$ supersymmetry have the following transformations under $H$:
\begin{equation}
Q^i_\alpha \;\; (\frac{1}{2},0,\frac{1}{2})^1,
\;\;\;\;\;\;\;\;\;\;\;
\overline Q_{i \dot\alpha } \;\;
 (0,\frac{1}{2},\frac{1}{2})^{-1},
\label{loli}
\end{equation}
where the superindex denotes the $U(1)_{\cal R}$ charge and the numbers
within parentheses the representations under each of the factors in
$SU(2)_L\otimes SU(2)_R \otimes SU(2)_I$.

The twist consists of considering as the rotation group the group
${\cal K}' = SU(2)_L'\otimes SU(2)_R$ where $SU(2)_L'$ is the diagonal subgroup
of $SU(2)_L\otimes SU(2)_I$. This implies that the isospin index $i$ becomes
a spinorial index $\alpha$: $Q^i_\alpha \rightarrow Q^\beta_\alpha$
and $\overline Q_{i\dot\beta } \rightarrow G_{\alpha\dot\beta}$. Precisely the
trace of $Q^\beta_\alpha$ is chosen as the generator of the scalar symmetry:
$Q = Q^\alpha_\alpha$. Under the new global group $H'={\cal K}'\otimes
U(1)_{\cal R}$, the symmetry generators transform as:
\begin{equation}
G_{\alpha\dot\beta} \;\; (\frac{1}{2},\frac{1}{2})^{-1},
\;\;\;\;\;\;\;\;\;\;
Q_{(\alpha\beta)} \;\; (1,0)^1,
\;\;\;\;\;\;\;\;\;\;
Q \;\; (0,0)^1.
\label{olga}
\end{equation}

Once the scalar symmetry is found we must study if, as stated in
(\ref{angela}), the energy-momentum tensor is exact,  \ie, if it can be written
as the transformation of some quantity under $Q$. The $N=2$ supersymmetry
algebra gives a  necessary condition for this to hold. Notice that, after
the twisting, such an algebra becomes:
\begin{equation}
\{Q^i_\alpha, \overline Q_{j\dot\beta} \} = \delta^i_j P_{\alpha\dot\beta}
\longrightarrow
\{ Q , G_{\alpha\beta} \} = P_{\alpha\dot\beta},
\label{conchi}
\end{equation}
where $P_{\alpha\dot\beta}$ is the momentum operator of the theory.
Certainly (\ref{conchi}) is only a necessary condition for the theory being
topological. However, up to date, for all the $N=2$ supersymmetric models whose
twisting has been studied, the relation on the right hand side of
(\ref{conchi}) has become valid for the whole energy-momentum tensor.

In ${\RR}^4$ the original and the twisted theory are equivalent. However, for
arbitrary manifolds  they are certainly different due to the fact that their
energy-momentum tensors are different. The twisting changes the spin quantum
numbers of the fields entering the theory and therefore their couplings to the
metric on $M$ are different.

As in all theories of Witten type the observables
are obtained from the $Q$-cohomology of the fields of the
theory. It is rather straightforward to prove that if one of the operators
entering (\ref{sanse}) is $Q$-exact the corresponding \vev\ vanishes.
In most of the cases the $Q$-cohomology which must be studied is an equivariant
cohomology.  Certainly this study must be done among the operators which are
invariant under $Q^2$. For example, for Yang-Mills theory,  the $N=2$
supersymmetry algebra closes up to gauge transformations (Wess-Zumino gauge).
This implies that  $Q^2$ is a gauge transformation and therefore the $Q$
cohomology must be studied on gauge invariant operators.

\vskip0.3cm
\noindent{\sc 3.2 BRST approach}
\vskip0.1cm

\noindent
Soon after Donaldson-Witten theory was formulated \cite{tqft},
the theory was reobtained by another method based on the application
of BRST gauge-fixing techniques \cite{lape}. This approach automatically
generates a scalar symmetry and an energy-momentum tensor which is $Q$-exact.
The basic ingredients of this approach are a set of basic fields, $\phi_i$, and
a set of basic equations $s^{(\alpha)}(\phi_i) = 0$. The main idea is to assume
that the theory possesses a symmetry which corresponds to arbitrary variations
of the basic fields and then BRST gauge-fix this symmetry using as gauge fixing
function the basic equations $s^{(\alpha)}(\phi_i) = 0$. As classical action
one
takes zero and therefore the construction automatically leads to an action
which is $Q$-exact. The $Q$-exactness of the energy-momentum follows if the
$Q$-transformations commute with the variations respect to the metric
$g_{\mu\nu}$. In the gauge-fixing process new fields are generated: ghost,
antighost and auxiliary fields. As in the previous case the observables are
obtained studying the corresponding equivariant $Q$-cohomology.

\vskip0.3cm
\noindent{\sc 3.3 Mathai-Quillen formalism}
\vskip0.1cm

\noindent
This formalism is the most geometrical one among all the approaches
leading  to the construction of TQFT. It can be applied to any
Witten type theory. Based on the work by Mathai and Quillen \cite{mathai} it
was
first implemented in the framework of TQFT by Atiyah and Jeffrey
\cite{jeffrey}.  The basic idea
behind this formalism is the extension to the  infinite-dimensional case of
ordinary finite-dimensional geometrical constructions. Soon after  the
formulation of the first TQFTs it became clear that the partition function
associated to most of these theories corresponds to the Euler class of
certain vector bundle related to the space of solutions of the basic
equations of
the theory (moduli space). In the finite-dimensional case there are
many forms to
obtain the Euler class. The Mathai-Quillen formalism basically consists of the
generalization of one of these forms to the infinite-dimensional case.

Let us consider a vector bundle $E$ of dimension $2m$ on a manifold $M$. This
bundle has its Euler class in $H^{2m}(M)$. If ${\hbox{\rm dim}} M =2m$ and $M$
is compact and orientable, the integration of one representative of the Euler
class, for example for the tangent bundle to $M$ the one supplied by the
Gauss-Bonet theorem, $\omega$, leads to the Euler number,
\begin{equation}
\varepsilon(E)=\int_M\omega.
\label{lola}
\end{equation}
The equations that
one obtains for the partition function of a TQFT
have a similar structure. However, such a formula is hard to implement in this
context because it lacks information on the main ingredient of a TQFT of
Witten type: the moduli space of solutions of its basic equations. Fortunately,
a most general form of (\ref{lola}) based on the Thom class is available.
Furthermore, this formula, besides a dependence on a connection on the vector
bundle $E$, it also depends on a section which certainly opens a room to
introduce the basic equations. This formula has the form,
\begin{equation}
\varepsilon(E)=\int_M\omega_s = \int_M s^* U,
\label{teresa}
\end{equation}
where $U$ is a representative of the Thom class. The generalization of this
expression to the infinite-dimensional case leads to the Mathai-Quillen
formalism. We will not describe here the details of this formalism. We refer
the reader to  \cite{moore,vafa,blau,matilde}  where  excellent
reviews of this approach are presented.

\vskip0.3cm
\noindent{\sc 3.4 Schwarz type theories}
\vskip0.1cm

\noindent
The previous methods considered refer to the case of Witten type
theories. In the case of Schwarz type theories one must first
construct an action
which is independent of the metric $g_{\mu\nu}$. The method is best illustrated
by considering an example. Let us take into consideration the most
interesting of
this type of theories: Chern-Simons gauge
theory \cite{csgt}. This is a three-dimensional theory whose action is the
integral of the Chern-Simons form associated to a gauge connection $A$
corresponding to a group $G$:
\begin{equation}
S_{\CS} (A) = \int_M \tr (A\wedge d A + \frac{2}{3} A\wedge
A\wedge A).
\label{valery}
\end{equation}
Observables must be constructed out of operators which do not contain the
metric
$g_{\mu\nu}$. In gauge invariant theories, as it is the case, one must also
demand invariance under gauge transformations for these operators. An important
set of the observables in Chern-Simons gauge theory is constituted by the trace
of the holonomy of the gauge connection $A$ in some representation $R$ along
a 1-cycle $\gamma$, \ie, the Wilson line:
\begin{equation}
\tr_R \big( {\hbox{\rm Hol}}_\gamma (A) \big) =
\tr_R {\hbox{\rm P}} \exp \int_\gamma A.
\label{silvie}
\end{equation}

\section{Chern-Simons gauge theory}

We have already introduced Chern-Simons gauge theory in the previous
subsection. This theory has a tremendous importance because the vevs of
its observables
are related to knot and link invariants of polynomial type as the Jones
polynomial and its generalizations (HOMFLY, Kauffman, Akutsu-Wadati, etc.).

The data in Chern-Simons gauge theory are the following: a differentiable
three-manifold $M$ which I will take  to be compact, a gauge group $G$, which
will be taken  simple and compact, and an integer parameter $k$. Once these
are chosen the observables are labeled by representations $R_i$ and embeddings
$\gamma_i$ of $S^1$ into $M$. They lead to the following vevs:
\begin{equation}
\langle \tr_{R_1} {\hbox{\rm P}} \ex^{\int_{\gamma_1} A}
        \dots
        \tr_{R_n} {\hbox{\rm P}} \ex^{\int_{\gamma_n} A} \rangle
        \nonumber \\
= \int [DA] \tr_{R_1} {\hbox{\rm P}} \ex^{\int_{\gamma_1} A}
        \dots
        \tr_{R_n} {\hbox{\rm P}} \ex^{\int_{\gamma_n} A}
        \ex^{\frac{i k}{4\pi} S_{\CSs} (A) }.
\label{encarna}
\end{equation}

The reason for $k$ being integer-valued is gauge invariance. Under a gauge
transformation $A\rightarrow g^{-1} d g + g^{-1} A g$ where $g : M\rightarrow
G$
the Wilson lines are invariant but the action of the theory transform as:
\begin{equation}
\frac{i k}{4\pi} S_{\CS} (A) \longrightarrow
\frac{i k}{4\pi} S_{\CS} (A) - \frac{ik}{12\pi}\int_M
\tr(g^{-1} d g\wedge g^{-1} d g\wedge g^{-1} d g).
\label{francisca}
\end{equation}
The second term in this equation has the form $-4i\pi y k w(g)$ where
$w(g)$ is an integer, the winding number of the map $g$, and $y$ is the Dynkin
index of the fundamental representation of $G$ which is half-integer. If $k$ is
an integer, the exponential of $\frac{i k}{4\pi} S_{\CS} (A)$ is gauge
invariant.

The non-perturbative analysis of the theory shows that the invariants
associated to the observables (\ref{encarna}) are knot and link invariants with
the same properties as the Jones polynomial and its generalizations. If one
considers $M=S^3$, $G=SU(2)$, and takes all the Wilson lines entering
(\ref{encarna})  in the fundamental representation, the non-perturbative
analysis proves that the vevs associated to three links
whose only difference is in an overcrossing, in an undercrossing
or in no-crossing,
satisfy the following relation:

\setlength{\unitlength}{5pt}
\begin{picture}(50,10)(-13,0)

\put(4,4.5){$q^{-1}$}
\put(7,2){\vector(1,1){6}}
\put(13,2){\line(-1,1){2.5}}
\put(9.5,5.5){\vector(-1,1){2.5}}
\put(15,4.5){$-$}
\put(17,4.5){$q$}
\put(19,2){\line(1,1){2.5}}
\put(25,2){\vector(-1,1){6}}
\put(22.5,5.5){\vector(1,1){2.5}}
\put(27,4.5){$=$}
\put(30,4.5){$(q^{\frac{1}{2}}-q^{-{\frac{1}{2}}})$}
\put(42,2){\vector(0,1){6}}
\put(45,2){\vector(0,1){6}}

\end{picture}

\noindent
where $q=\exp(2\pi i / (2yk + g^\vee))$ being $g^\vee$ the dual Coxeter number
of the group $G$. These are precisely the skein rules which define the Jones
polynomial. The great advantage of Chern-Simons gauge theory is
that it allows to
generalize very simply these invariants to other groups and other
representations. The HOMFLY \cite{homfly} and the Kauffman \cite{kauffman}
polynomials are obtained after considering the fundamental representation of
the groups $SU(N)$ and $SO(N)$, respectively. The Akutsu-Wadati \cite{aku} or
colored Jones polynomial is obtained  considering the group $SU(2)$ with Wilson
lines in different representations. Other non-perturbative methods have allowed
to obtain these invariants for classes of knots and links as, for example,
torus knots and links \cite{torus}. Methods for general computations of these
invariants have been proposed in \cite{martin} and \cite{kaul}.


{}From the point of view of perturbation theory, Chern-Simons gauge theory has
led
to important results. Furthermore, this theory has been studied in both, the
Hamiltonian (non-covariant) and the Lagrangian (covariant) approaches providing
a variety of interesting results. Besides, in each of these cases it is
possible to analyze the theory in different gauges obtaining in this way
remarkable relations.

The perturbative formulation of a quantum field theory is contained in its
Feynman rules. In the case of Chern-Simons gauge theory these
Feynman rules lead to the following general power series expansion for the
vev of a Wilson line \cite{alla}:
\begin{equation}
\frac{\langle \tr_R {\hbox{\rm P}} \exp \int_\gamma A \rangle}
     {\langle 1 \rangle} =
d(R)\sum_{i=0}^\infty \sum_{j=1}^{d_i} \alpha_{ij}(\gamma) r_{ij}(R) x^i,
\label{mcarmen}
\end{equation}
where $r_{ij}(R)$ (group factors) contain all the dependence on the
representation and the group which has been chosen, and $\alpha_{ij}(\gamma)$
(geometric factors)  all the dependence on the path $\gamma$ and the
three-dimensional manifold where this path is embedded. In (\ref{mcarmen})
$i$ corresponds to the order in perturbation theory and $x=i\pi/yk$.
The integers $d_i$ denote the number of independent group factors.
Previous analysis of the lowest order of the series (\ref{mcarmen})
can be found in \cite{gmm} and \cite{natan}.

The first problem that one has to face to study the expansion (\ref{mcarmen})
is the classification of the independent group factors. The Feynman rules of
the theory are such that this problem can be stated in terms of trivalent
graphs. Let $\Gamma$ be the set of all diagrams containing a circle and a
trivalent graph. The set of independent group factors is isomorphic to $\Gamma$
modulo the AS, IHX and STU relations given in \cite{barnatan}. Let us denote by
${\cal B}$ the resulting set of diagrams. This set has a natural grading
($1/2$ the number of trivalent vertices) which precisely coincides with the
order in perturbation theory (the power of $x$ in (\ref{mcarmen})).
The number of elements of ${\cal B}$ at order $i$ are therefore the quantities
$d_i$ appearing in (\ref{mcarmen}). These numbers are known only up to order
$i=9$ \cite{barnatan} and are given in Table \ref{latablados}.
\begin{table}[htbp]
\centering
\begin{tabular}{|c|c|c|c|c|c|c|c|c|c|}  \hline
 $i$ & 1 & 2 & 3 & 4 & 5 & 6 & 7 & 8 & 9 \\  \hline
 $d_i$ & 0 & 1 & 1 & 3 & 4 & 9 & 14 & 27 & 44 \\  \hline
\end{tabular}
\caption{Number of independent group factors}
\label{latablados}
\end{table}

In the power series expansion (\ref{mcarmen}) the group factor is independent
of the knot which has been chosen. All the dependence on the knot is in the
geometric factors. Since the total sum is a topological invariant and at
each order the geometric factors multiply quantities which are independent,
the geometric factors are knot invariants. Chern-Simons gauge theory supplies
an infinite set of knot invariants with a natural grading given by the order
in perturbation theory. Furthermore, it provides an explicit
integral representation for each of them. At order $i=2$, for the case in which
$M={\RR}^3$, one has \cite{gmm}:
\begin{eqnarray}
\alpha_{21}(\gamma)&=&\frac{1}{4}\oint dx^\mu \int^x d y^\nu \int^y dz^\rho
\int^z dw^\sigma \Delta_{\mu\rho}(x-z) \Delta_{\nu\sigma}(y-w) \nonumber \\
&-&\frac{1}{16}\oint dx^\mu \int^x d y^\nu \int^y d z^\rho
\int_{{\RRs}^3} d^3\omega v_{\mu\nu\rho}(x,y,z;\omega),
\label{sara}
\end{eqnarray}
where,
\begin{equation}
\Delta_{\mu\nu}(x-y) = \frac{1}{\pi}\epsilon_{\mu\rho\nu}
\frac{(x-y)^\rho}{|x-y|^3}, \;\;\;
v_{\mu\nu\rho}(x,y,z;\omega)=\Delta_{\mu\sigma_1}(x-\omega)
\Delta_{\nu\sigma_2}(y-\omega)\Delta_{\rho\sigma_3}(z-\omega)
\epsilon^{\sigma_1 \sigma_2 \sigma_3}.
\label{adelaida}
\end{equation}

The knot invariants $\alpha_{ij}$ turn out to be Vassiliev invariants or of
finite type. A knot invariant is of type $n$ if after defining invariants for
singular knots via the relation,

\begin{picture}(50,10)(-13,0)

\put(10,2){\vector(1,1){6}}
\put(16,2){\vector(-1,1){6}}
\put(18,4.5){$=$}
\put(21,2){\vector(1,1){6}}
\put(13,5){\circle*{0.8}}
\put(27,2){\line(-1,1){2.5}}
\put(23.5,5.5){\vector(-1,1){2.5}}
\put(29,4.5){$-$}
\put(32,2){\line(1,1){2.5}}
\put(35.5,5.5){\vector(1,1){2.5}}
\put(38,2){\vector(-1,1){6}}

\end{picture}

\noindent
one finds that it vanishes for all knots containing $n$ singularities.
Birman and Lin showed \cite{bilin} that for any polynomial invariant
arising from
Chern-Simons gauge theory with semisimple gauge group, after expressing
$q$ as,
\begin{equation}
q=\exp \Big( \frac{2\pi i}{2yk+g^\vee}\Big) = \exp(x),
\label{irene}
\end{equation}
and expanding in power series of $x$, the coefficient of $x^n$ is an invariant
of type $n$. This implies that the invariants $\alpha_{ij}$ are Vassiliev
invariants of type $n$. Using the results by Bar-Natan in \cite{barnatan}
based on the integral representation for Vassiliev invariants provided by
Kontsevich \cite{kont} one can show that the quantities $\alpha_{ij}$
supplied by
Chern-Simons gauge theory constitute a complete set of Vassiliev invariants.

As we have been discussing in this section, Chern-Simons gauge
theory provides a covariant integral representation for Vassiliev invariants
\cite{alla}. This representation has been considered recently by Bott and
Taubes \cite{botttaubes} from a different perspective. Kontsevich's
representation for Vassiliev invariants \cite{kont} is not covariant. It can be
obtained from Chern-Simons gauge theory  by analyzing the theory in the
Hamiltonian formalism \cite{mateos}. Recent work shows that it can also be
obtained from a Lagrangian analysis of the theory in a non-covariant gauge
\cite{cotta}.

Using the perturbative expansion (\ref{mcarmen}) Vassiliev invariants have been
computed up to order six for all prime knots up to six crossings
\cite{alla} and for arbitrary torus knots \cite{torusknots}. These results have
been obtained using the fact that Chern-Simons gauge theory, as any of the
TQFTs
studied up to date, have a perturbative series expansion which is exact, \ie,
is
equivalent to its non-perturbative counterpart.

\section{Donaldson-Witten theory and its generalizations}

As explained in the introduction, the twisted version of $N=2$ supersymmetric
Yang-Mills in four dimensions provides a TQFT
\cite{tqft} whose
observables are related to Donaldson invariants. This theory  has been
extended recently \cite{nabm} to a wide range of moduli spaces
whose associated invariants turn out to satisfy unexpected relations.

Let $M$ be a compact oriented four-dimensional manifold endowed with a
metric $g_{\mu\nu}$, and let us consider on it a principal fibre bundle $P$
with
group
$G$ which will be assumed to be simple, compact and connected. Let $E$ be the
vector bundle associated to $P$ via the adjoint representation, and let ${\cal
A}$ be the space of $G$-connections on $E$. A connection in ${\cal A}$ will be
denoted by $A$ and its corresponding covariant derivative and self-dual part of
its curvature by $D_\mu$ and $F^+$, respectively. Let us introduce the
following set of fields:
\begin{equation}
\chi_{\mu\nu}, \; G_{\mu\nu} \; \in \; \Omega^{2,+}(M,{\mbox{\bf g}}),
\;\;\;\;\;
\psi_\mu \; \in \; \Omega^1(M,{\mbox{\bf g}}),
\;\;\;\;\;
\eta, \; \lambda, \; \phi \; \in \; \Omega^{0}(M,{\mbox{\bf g}}).
\label{ines}
\end{equation}
In (\ref{ines}) ${\mbox{\bf g}}$ denotes the Lie algebra associated to $G$.
The action of the theory has the form:
\begin{eqnarray}
S = \int_M d^4 x\,\sqrt{g}&&\tr\Big({F^+}^2-G^2-i\chi^{\mu\nu} D_\mu \psi_\nu
+i\eta D_\mu \psi^\mu
+\frac{1}{4} \phi \{ \chi_{\mu\nu},\chi^{\mu\nu} \} \nonumber \\
&&+\frac{i}{4} \lambda \{\psi_\mu,\psi^\mu\} - \lambda D_\mu D^\mu \phi
+ \frac{i}{2}\phi\{\eta,\eta\}+\frac{1}{8}[\lambda,\phi]^2 \Big)
\label{adela}
\end{eqnarray}

The scalar symmetry which characterizes the theory has the form:
\begin{eqnarray}
\delta A_\mu &= \psi_\mu,  \mbox{\hskip2cm} \delta \chi_{\mu\nu} &= G_{\mu\nu},
\nonumber \\ \delta \psi &= d_A\phi, \mbox{\hskip1.5cm}
\delta G_{\mu\nu} &= i[\chi_{\mu\nu},\phi], \nonumber \\
\delta \phi &= 0, \mbox{\hskip2.3cm}
\delta\lambda &= \eta,
\mbox{\hskip2cm} \delta\eta = i [\lambda,\phi].
\label{adelados}
\end{eqnarray}
These transformations close up to a gauge transformation and the action
(\ref{adela}) is $\delta$-exact. This implies, on the one hand,
that the action is
invariant under (\ref{adelados}); on the other hand that its energy-momentum
tensor is $\delta$-exact and therefore that the theory is topological.
The $\delta$-cohomology associated to (\ref{adelados}) was studied in
\cite{tqft}. For the case $G=SU(2)$ the resulting observables are based on the
following operators:
\begin{eqnarray}
W_0 &= \frac{1}{2} \tr (\phi^2)  \mbox{\hskip4.5cm} W_1 &=
\tr(\phi\wedge\psi), \nonumber \\
W_2 &= \tr(\frac{1}{2}\psi\wedge\psi+i\phi\wedge F), \mbox{\hskip2cm}
W_3 &= i\tr(\psi\wedge F).
\label{raquel}
\end{eqnarray}
These operators satisfy the descent equations, $\delta W_i = d W_{i-1}$, which
allow to define the following observables:
\begin{equation}
{\cal O}^{(k)} = \int_{\gamma_k} W_k,
\label{sonia}
\end{equation}
where $\gamma_k \in H_k(M)$. The descent equations imply that they are
$\delta$-invariant.

The functional integral corresponding to the topological invariants  of the
theory has the form:
\begin{equation}
\langle {\cal O}^{(k_1)} {\cal O}^{(k_2)} \cdots
{\cal O}^{(k_p)} \rangle =
\int   {\cal O}^{(k_1)} {\cal O}^{(k_2)} \cdots
{\cal O}^{(k_p)} \exp ({-{S}/{g}}),
\label{clara}
\end{equation}
where the integration has to be understood on the space of field configurations
modulo gauge transformations and $g$ is a coupling constant. Standard arguments
\cite{tqft} show that due to the $\delta$-exactness of the theory the
quantities obtained in (\ref{clara}) are independent of $g$. This implies that
the observables of the theory can be obtained either in the limit
$g\rightarrow 0$, where perturbative methods apply, or in the limit
$g\rightarrow \infty$ where one is forced to use non-perturbative ones. The
crucial point is to observe that the $\delta$-exactness of the
action implies, at
least formally, that in either case the quantities obtained must be the same.

The previous argument for $g\rightarrow 0$ implies that the semiclassical
approximation of the theory is exact. In this limit the contributions to the
functional integral are dominated by the  field configurations which minimize
$S$. Let us assume that in the situation under consideration there are only
irreducible connections. In this case the contributions from the even
part of the
action are given by the solutions of the equation $F^+=0$, \ie, by instanton
configurations. Being the connection irreducible there are no non-trivial
solution to the
classical equations for the fields $\lambda$ and $\phi$.

{}From the odd part of
the action one finds that the only contributions come from the solutions to the
equations,
\begin{equation}
(D_\mu\psi_\nu)^+=0, \;\;\;\;\;\;\;\; D_\mu\psi^\mu=0,
\label{rosana}
\end{equation}
which are precisely
the ones that define the tangent space to the space of instanton
configurations.
The number of independent solutions of these equations determine the dimension
of the instanton moduli space ${\cal M}$. These dimensions can be computed with
the help of index theorems. For $SU(2)$ one has:
$d_{\cal M} = 8k-3(\chi+\sigma)/2$, where $\chi$ is the Euler number and
$\sigma$ the signature of the manifold $M$, and $k$ is the instanton number.

The fundamental contribution to the functional integral (\ref{clara})
is given by
the elements of ${\cal M}$ and by the zero-modes of the solutions to
(\ref{rosana}). Once these have been obtained they must be introduced in the
action and an expansion up to quadratic terms in non-zero modes must be
performed. These are the only relevant terms in the limit $g\rightarrow 0$. The
resulting gaussian integrations then must be performed. Due to the presence of
the
$\delta$ symmetry these come in quotiens whose value is $\pm 1$. The functional
integral (\ref{clara}) takes the form\footnote{This equation is a simplified
form of the true one. A more detailed analysis shows that the
field $\phi$ has to be
replaced by its vev.}:
\begin{equation}
\langle {\cal
O}^{(k_1)} {\cal O}^{(k_2)} \cdots {\cal O}^{(k_p)} \rangle =
\int_{\cal M} da_1\cdots da_{d_{\cal M}} d\psi_1 \cdots d\psi_{d_{\cal M}}
{\cal O}^{(k_1)} {\cal O}^{(k_2)} \cdots {\cal O}^{(k_p)}
(-1)^{f(a_1,\dots,a_{d_{\cal M}})},
\label{clarados}
\end{equation}
where $f(a_1,\dots,a_{d_{\cal M}})=0,1$. The integration over the
odd modes leads
to a selection rule for the product of observables. This selection rule is
better expressed making use of the quantum numbers of the fields
associated to the $U(1)_{\cal R}$ symmetry inherited from $N=2$
supersymmetry. These quantum numbers are usually called ghost numbers due to
its origin in the BRST approach to TQFT. For the operators  in
(\ref{sonia}) one has: $U({\cal O}^{(k)})=4-k$, and  the selection rule
can be written as $d_{\cal M}=\sum_{i=1}^p U({\cal O}^{(k_i)})$.

In the case in which $d_{\cal M}=0$, the only observable is the partition
function which takes the form $\langle 1 \rangle = \sum_i (-1)^{f_i}$, where
the sum is over isolated instantons. In general, the integration of the
zero-modes in  (\ref{clarados}) leads to an antisymmetrization in such a way
that one ends with the integration of a $d_{\cal M}$-form on ${\cal M}$.
The resulting real number is a topological invariant.
Notice that in the process a map $H_k(M) \longrightarrow H^k({\cal M})$
has been constructed. The vevs of the theory provide
polynomials in $H_{k_1}(M) \times H_{k_2}(M) \times \cdots \times H_{k_p}(M)$
which are precisely the Donaldson polynomials.

The study of Donaldson-Witten theory from a perturbative point of view
has allowed to show that the vevs of the observables of
this theory are related to Donaldson invariants. However, it does not provide a
new method to compute them since the integral functional leads to an
integration
over the moduli space of instantons, which is precisely the step where the
hardest problems to compute these invariants appear. From a quantum field
theory we have still the possibility of studying the form of these observables
from a non-perturbative point of view, \ie, their study in the limit
$g\rightarrow \infty$. This line of research seemed difficult to implement
until very recently. However, in  1994,
after the work by Seiberg and Witten \cite{sw}, important progress was
made in the knowledge of the non-perturbative structure of $N=2$
supersymmetric Yang-Mills theories. These results have been
immediately applied to the twisted theory leading to explicit expressions for
the topological invariants in a variety of situations \cite{abm}. But perhaps
the most important result achieved in this context is the existence of a
relation between the moduli space of instantons and other moduli spaces such
as the moduli space of abelian monopoles.

$N=2$ supersymmetric Yang-Mills theory is asymptotically free. This means that
the effective coupling constant becomes small at large energies. The
perturbative methods which have been used are therefore valid at these
energies.  At low energies, however,  those methods are not valid and one must
use non-perturbative methods. From the work in \cite{sw} follows that at low
energies the theory behaves as an abelian gauge theory. The
effective theory is parametrized  by a complex variable $u$ which labels the
vacuum structure of the theory. The most important feature of the effective
theory is that there are points in the $u$-complex plane where monopoles and
dyons become massless. These points are singular points and there are two for
$SU(2)$, $u=\pm\Lambda^2$, where $\Lambda$ is the dynamically generated scale
of the theory.  For the point $u=\Lambda^2$ the theory consists of an $N=2$
supersymmetric abelian gauge theory coupled to a massless monopole, while for
$u=-\Lambda^2$ it is coupled to a dyon. It is important to remark that the
${\cal R}$ symmetry of the initial theory is broken to a ${\ZZ}_2$ symmetry.
{}From the breaking process one can extract the transformation which
relates the behavior of the  theory around one singularity in terms of
its behavior around the other. This is very important because it is not
possible
to describe both with a local theory.

The theory around a massless monopole is an $N=2$
supersymmetric abelian gauge theory coupled to a massless $N=2$
hypermultiplet. This theory has its associated twisted version which is the
one that must correspond to the low energy behavior of Donaldson-Witten theory
or large coupling limit, $g\rightarrow \infty$. This version has been
constructed in \cite{abmono} using the Mathai-Quillen formalism.
The structure of
this theory is similar to the one of Donaldson-Witten theory. The resulting
action is $\delta$-exact and therefore one can study the theory in the weak
coupling limit, which, being abelian, corresponds to the low energy limit. The
main contribution to the functional integral from the even part of the action
is
given by the solutions to the equations, \begin{equation}
F_{\mu\nu}^+ + \frac{i}{2} \overline M \gamma_{\mu\nu}^+ M =0,
\;\;\;\;\; \gamma^\mu D_\mu M = 0,
\label{francesca}
\end{equation}
where $M$ is a Weyl spinor of  $U(1)$ charge one. Let us denote by $L$ the
complex line bundle to which the abelian connection $A$ is associated. The
Weyl spinor $M$ is a section  of $S^+\otimes L$, being $S^+$ the positive
chirality spin bundle. Of course, we are assuming that our manifold is a spin
manifold . If this were not the case a similar analysis can be carried out
introducing a $Spin_c$ structure. The equations (\ref{francesca}) are known as
monopole equations \cite{abm}.

We have studied the twisted theory around one of the singular points. However,
vevs must be computed integrating over the possible values of the
parameter $u$. The crucial point which implies calculability for the twisted
theory is that it is possible to argue that in a wide variety of situations
the only contributions come from the singular points. For a generic value of
$u$ the possible contributions could occur if there are abelian instantons (for
a generic $u$ there are no other massless modes). Manifolds with $b^+_2>1$ do
not possess abelian instantons and therefore for this case one just has to sum
over singular points.

Let us describe first the contribution from $u=\Lambda^2$. The selection rule
corresponding to the TQFT of the twisted effective
theory is similar to the one for the partition function of Donaldson-Witten
theory. The dimension of the moduli space of monopoles ${\tilde{\cal M}}$ is
$d_{\tilde{\cal M}}= -(2\chi+3\sigma)/4+x^2/4$ where $x=-2c_1(L)$. The
selection
rule
$d_{\tilde{\cal M}}=0$ implies $x^2=2\chi+3\sigma$. Let us denote by $n_x$ the
resulting partition function. The classes $x$ such that $n_x\neq 0$ are called
basic classes and the quantities $n_x$, that being the partition function of a
TQFT are topological invariants, are called
Seiberg-Witten invariants. Reviews on these invariants have appeared recently
\cite{matilde,yang}.

Instead of considering products of observables as before, let us consider their
generating function. We will restrict ourselves to the case in which the
four-manifold $M$ is simply connected so that $H_1(M)=H_3(M)=0$. In this
situation the only observables are the ones associated to  $\Sigma_a \in
H_2(M)$
and to points
$x\in M$. If we denote by $I(\Sigma_a)$ the operator $\int_{\Sigma_a} W^{(2)}$
and by
${\cal O}$ the one to $W^{(0)}(x)$ we must study the following vev:
\begin{equation}
\langle \exp \big( \sum_a \alpha_a I(\Sigma_a) + \mu {\cal O}\big) \rangle,
\label{karen}
\end{equation}
where $\alpha_a$ and $\mu$ are parameters and $a$ runs over a basis in
$H_2(M)$.
At the singular point corresponding to the massless magnetic monopole
the vev of this observable takes the form:
\begin{equation}
\exp \big( \gamma v^2 + \mu \langle {\cal O} \rangle \big)
\sum_x n_x \ex^{\langle V \rangle v\cdot x},
\label{monica}
\end{equation}
where $v^2=\sum_{a,b} \alpha_a \alpha_b \# (\Sigma_a \cap \Sigma_b)$,
$v\cdot x = \sum_a \alpha_a \int_{\Sigma_a} x$, and $\gamma$, $\langle {\cal O}
\rangle $ and $\langle  V \rangle $ are universal constants, independent of the
manifold
$M$. To motivate the origin of (\ref{monica}) we just will indicate that $v^2$
is generic \cite{wijmp}, and that in the operator $I(\Sigma_a) =
\int_{\Sigma_a}
(i\phi F + \psi \wedge \psi /2 )$ only the first term contributes. It has the
form $v\cdot x$ since at the singular point $\phi$ takes a constant value.

Once the contribution from the first singular point has been computed, the one
from the other is easily obtained using the ${\ZZ}_2$ symmetry. In this case
this symmetry is:
\begin{equation}
{\cal O} \longrightarrow -{\cal O},
\;\;\;\;\;\;\;
I(\Sigma_a) \longrightarrow  -i I(\Sigma_a).
\label{denise}
\end{equation}
There is in addition a factor due to the global anomaly of the
${\ZZ}_2$ symmetry. It has the form $i^\Delta$ where $\Delta =
(\chi+\sigma)/4$. Finally the vev (\ref{karen}) takes the
form:
\begin{equation}
c\Big(
\exp \big( \gamma v^2 + \mu \langle {\cal O} \rangle \big)
\sum_x n_x \ex^{\langle V \rangle v\cdot x} +
i^\Delta \exp \big( -\gamma v^2 - \mu \langle {\cal O} \rangle \big)
\sum_x n_x \ex^{-i\langle V \rangle v\cdot x}
\Big).
\label{tania}
\end{equation}
The constants $c$, $\gamma$, $\langle {\cal O} \rangle$ and $\langle V \rangle$
can be calculated comparing to the expression obtained by Kronheimer and
Mrowka \cite{km}:
\begin{equation}
c = 2^{1+(7\chi + 11\sigma)/4},
\;\;\;
\gamma=1,
\;\;\;
\langle {\cal O} \rangle =2,
\;\;\;
\langle V \rangle =1.
\label{louise}
\end{equation}

Similar methods can be used for other groups different than $SU(2)$. In
general, the number of singularities is bigger. For example, for $SU(N)$ one
has $N$ singularities. Recently, models which generalize Donaldson-Witten
theory
have been proposed. These theories are based on non-abelian monopoles. They can
be understood as a twisted version of $N=2$ supersymmetric Yang-Mills coupled
to
$N=2$ supersymmetric matter \cite{park} (see also \cite{rocek,top,ans}). They
have been proposed using the Mathai-Quillen formalism in \cite{nabm}. The fact
that they can be thought as twisted versions of $N=2$ supersymmetric theories
allows to use the recent results on the non-perturbative behavior of these
theories. The basic equations for $SU(N)$ non-abelian monopoles are:
\begin{equation}
F^{+ij}_{\mu\nu}+\frac{1}{2}\big(\overline M^j_{(\mu} M^i_{\nu)}
-\frac{\delta^{ij}}{N} \overline M^k_{(\mu} M^k_{\nu)} \big) = 0,
\;\;\;\;\;\;
\gamma^\mu D_\mu M = 0,
\label{telma}
\end{equation}
where $M^i \in \Gamma( M, S^+\otimes E)$, \ie, $M^i$ is a Weyl spinor in the
fundamental representation of $SU(N)$. As before in the weak coupling limit
there appear a  structure similar to the case of Donaldson-Witten theory. For
strong coupling one finds again an abelian theory \cite{sw} but this time
there are three singularities related by a ${\ZZ}_3$ symmetry. The most
important aspect of this analysis is that the vevs of observables can be
expressed in terms of the Seiberg-Witten invariants $n_x$ as in the previous
case \cite{last}. A new study of this model  has
appeared recently \cite{parkdos}.

\section{Concluding Remarks}

The generalization of Donaldson-Witten theory opens the study of an enormous
set
of TQFTs. Each of these new theories can be labeled by a gauge group and a
representation (or several representations if there is more than one flavor).
One would like to know for which
of these theories the vevs of observables can be expressed in terms of
Seiberg-Witten invariants. Is this true for all of them, or Seiberg-Witten
invariants are just the first of a family of invariants which allow to express
all the vevs in terms of them?

A look at (\ref{tania}) reveals that almost all
the topological information of the vevs is contained in the Seiberg-Witten
invariants $n_x$. The rest are factors fixed by the group and representation
(or representations) chosen. This structure is
similar to the one found for Chern-Simons gauge theory in three dimensions (see
(\ref{mcarmen})). There the group factors do not contain information on the
topology of the three manifold, all the topological information is contained in
the geometric factors or Vassiliev invariants. In this sense there is a
parallelism between   Vassiliev and Seiberg-Witten invariants: both contain the
topological information of the manifold $M$ and are universal respect to the
group and representation (or representations) chosen.

\begin{center}
{\bf Acknowledgements.}
\end{center}

\vspace{4 mm}

I would like to thank M. Salgado and the organizers of the Workshop for
inviting
me to give this talk. I would like to thank also A.V. Ramallo and my
collaborators M. Alvarez, M. Mari\~no and E. P\'erez for very helpful
discussions on many aspects of the subjects treated in this talk.
This work was supported in part by DGICYT under grant
PB93-0344 and  by CICYT under grant AEN94-0928.

\newpage



\begin{thebibliography}{99}


\def\np{Nucl. Phys.}
\def\pl{Phys. Lett.} \def\pre{Phys. Rep.} \def\prl{Phys. Rev. Lett.}
\def\pr{Phys. Rev.} \def\ap{Ann. Phys.} \def\cmp{Comm. Math. Phys.}
\def\ijmp{Int. J. Mod. Phys.} \def\mpl{Mod. Phys. Lett.} \def\lmp{Lett.
Math. Phys.} \def\bams{Bull. AMS} \def\am{Ann. of Math.} \def\jpsc{J. Phys.
Soc. Jap.} \def\topo{Topology} \def\ijm{Int. J. Math.} \def\knot{Journal of
Knot Theory and Its Ramifications} \def\jmp{J. Math. Phys.} \def\jgp{J.
Geom. Phys.} \def\jdg{J. Diff. Geom.}
\def\plms{Proc. London Math. Soc.}
\def\mrl{Math. Res. Lett.}
\def\inma{Invent. Math.}
\def\tam{Trans. Am. Math. Soc.}



\bibitem{morse} E. Witten, {\sl\jdg} {\bf 17} (1982) 661.

\bibitem{floer} A. Floer, {\sl\bams} {\bf 16} (1987) 279.

\bibitem{atiyah} M. F. Atiyah, ``New invarianst of three and four dimensional
manifolds", in {\it The Mathematical Heritage of Herman Weyl}, {\sl Proc.
Symp. Pure Math.} {\bf 48}, {\sl American Math. Soc.} (1988) 285-299.

\bibitem{tqft} E. Witten, {\sl\cmp} {\bf 117} (1988) 353.

\bibitem{donald}S. K. Donaldson, {\sl\topo} {\bf 29} (1990) 257.

\bibitem{tsm} E. Witten, {\sl\cmp} {\bf 118} (1988) 411.

\bibitem{csgt} E. Witten,  {\sl\cmp} {\bf 121} (1989) 351.

\bibitem{gromov} M. Gromov,  {\sl\inma} {\bf 82} (1985) 307.

\bibitem{jones} V. F. R. Jones, {\sl bams} {\bf 12} (1985) 103;
{\sl\am} {\bf 126} (1987) 335.

\bibitem{sw} N. Seiberg and E. Witten, {\sl\np} {\bf B426} (1994) 19;
Erratum, {\sl\np} {\bf B430} (1994) 485; {\sl\np} {\bf B431} (1994) 484.

\bibitem{vass} V. A. Vassiliev, ``Cohomology of knot spaces", {\it Theory of
singularities and its applications}, {\sl Advances in Soviet Mathematics},
vol. 1, {\sl Americam Math. Soc.}, Providence, RI, 1990, 23-69.

\bibitem{abm} E. Witten, {\sl\mrl} {\bf 1} (1994) 769.

\bibitem{nabm} J. M. F. Labastida and M. Mari\~no, {\sl\np} {\bf B448} (1995)
373.

\bibitem{thompson} D. Birmingham, M. Blau, M. Rakowski and G. Thompson, {\sl
Phys. Rep.} {\bf 209} (1991) 129.

\bibitem{moore} S. Cordes, G. Moore and S. Rangoolam,
``Lectures on  2D Yang-Mills Theory, Equivariant Cohomology and
Topological Field Theory", hep-th/9411210, November 1994.

\bibitem{axio} M. Atiyah, ``The geometry and physics of knots",
 Cambridge University Press, 1990.

\bibitem{lape} J.M.F. Labastida and M. Pernici, {\sl \pl} {\bf B212} (1988) 56;
L. Baulieu and I.M. Singer, {\sl \np} {\bf B} (Proc. Suppl.) {\bf 5B} (1988)
12; R. Brooks, D. Montano and J. Sonnenschein, {\sl \pl} {\bf B214} (1988) 91.

\bibitem{mathai} V. Mathai and D. Quillen, {\sl Topology} {\bf 25} (1986)
85.

\bibitem{jeffrey} M. Atiyah and L. Jeffrey, {\sl J. Geom. Phys.} {\bf 7} (1990)
119.

\bibitem{vafa} C. Vafa and E. Witten, {\sl \np} {\bf B431} (1994) 3.

\bibitem{blau} M. Blau and G. Thompson,
{\sl\jmp} {\bf 36} (1995) 2192.

\bibitem{matilde} M. Marcolli, ``Notes on Seiberg-Witten gauge theory",
dg-ga/9509005, August 1995.

\bibitem{yang} J. V. Yang, ``Introduction to Seiberg-Witten invariants",
dg-ga/9508005, October 1995.


\bibitem{homfly} P. Freyd, D. Yetter, J. Hoste, W.B.R. Lickorish, K. Millet
and A. Ocneanu, {\sl\bams} {\bf 12} (1985) 239.

\bibitem{kauffman} L.H. Kauffman, {\sl\tam} {\bf 318} (1990) 417.

\bibitem{aku} Y. Akutsu and M. Wadati,
{\sl\jpsc} {\bf 56} (1987) 839; {\sl\jpsc} {\bf 56} (1987) 3039.

\bibitem{torus} J.M.F. Labastida and A.V. Ramallo, {\sl\pl} {\bf B227}
(1989) 92; J.M.F. Labastida, P.M. Llatas and A.V. Ramallo,
{\sl\np} {\bf B348} (1991) 651; J.M. Isidro, J.M.F. Labastida and A.V. Ramallo,
{\sl\np} {\bf B398} (1993) 187; J.M.F. Labastida and
M. Mari\~no, {\sl \ijmp} {\bf A10} (1995) 1045; J.M.F.
Labastida and E. P\'erez, ``A relation between the Kauffman and the HOMFLY
polynomials for torus knots", Santiago preprint, q-alg/9507031, May 1995.

\bibitem{martin} S. Martin, {\sl\np} {\bf B338} (1990) 244.

\bibitem{kaul} R.K. Kaul, T.R. Govindarajan, {\sl\np} {\bf B380} (1992) 293;
{\sl \np} {\bf B393} (1993) 392; {\sl\np} {\bf B402} (1993) 548.

\bibitem{alla} M. Alvarez and J.M.F.
Labastida, {\sl\np} {\bf B395} (1993) 198; {\sl\np} {\bf B433} (1995) 555;
Erratum, {\sl\np} {\bf B441} (1995) 403.

\bibitem{gmm} E. Guadagnini, M.
Martellini and M. Mintchev, {\sl\pl} {\bf B227} (1989) 111;
{\sl\pl} {\bf B228} (1989) 489; {\sl\np} {\bf B330} (1990) 575.

\bibitem{natan} D. Bar-Natan, ``Perturbative aspects of Chern-Simons
topological quantum field theory", Ph. D. Thesis, Princeton University,
1991.

\bibitem{barnatan} D. Bar-Natan, {\sl\topo} {\bf 34} (1995) 423.

\bibitem{kont} M. Kontsevich, {\sl Advances in Soviet Math.} {\bf 16}, Part 2
(1993) 137.

\bibitem{bilin} J.S. Birman and X.S. Lin, {\sl \inma} {\bf 111} (1993) 225;
J.S. Birman, {\sl\bams} {\bf 28} (1993) 253; X.S. Lin, Vertex models, quantum
groups and Vassiliev knot invariants, Columbia University preprint, 1991.

\bibitem{botttaubes} R. Bott and C. Taubes, {\sl\jmp} {\bf 35} (1994) 5247.

\bibitem{mateos} J. Mateos, {\sl J. Geom. Phys.} {\bf 15} (1994) 1.

\bibitem{cotta} A.S.Cattaneo, P.Cotta-Ramusino, A.Gamba and M.Martellini,
``The Donaldson-Witten Invariants in Pure QCD with Order and Disorder 't
  Hooft-like Operators", hep-th/9502110, February 1995.

\bibitem{torusknots} M. Alvarez and J.M.F. Labastida,
``Vassiliev invariants for torus knots", q-alg/9506009, June 1995.

\bibitem{abmono} J.M.F. Labastida and M. Mari\~no,
{\sl\pl} {\bf B351} (1995) 146.

\bibitem{wijmp} E. Witten, {\sl \jmp} {\bf 35} (1994) 5101.

\bibitem{park} S. Hyun, J. Park and J.S. Park,
``Topological QCD", hep-th/9503201.

\bibitem{rocek} A. Karlhede and M. Ro\v cec, {\sl\pl} {\bf B212} (1988) 51.

\bibitem{top} M. Alvarez
and J.M.F. Labastida, {\sl\pl} {\bf B315} (1993) 251;
{\sl\np} {\bf B437} (1995) 356.

\bibitem{ans} D. Anselmi and P. Fr\`e, {\sl\np} {\bf B392} (1993) 401;
{\sl\np} {\bf B404} (1993) 288; {\sl\np}
{\bf B416} (1994) 25; {\sl\pl} {\bf B347} (1995) 247.

\bibitem{km} P.B. Kronheimer and T.S. Mrowka, {\sl \bams} {\bf 30} (1994) 215.

\bibitem{last} J.M.F. Labastida and M. Mari\~no, ``Polynomial invariants
for $SU(2)$ monopoles", hep-th/9507140, July, 1995.

\bibitem{parkdos} S. Hyun, J. Park and J.S. Park,
``N=2 Supersymmetric QCD and Four Manifolds", hep-th/9508162, August 1995.




\end{thebibliography}
\end{document}